\documentstyle[aps,preprint,tighten,epsfig]{revtex}

\newcommand{\bea}{\begin{eqnarray}}
\newcommand{\eea}{\end{eqnarray}}
\newcommand{\ba}{\begin{array}}
\newcommand{\ea}{\end{array}}

\newcommand{\vp}{{\bar p}}

\newcommand{\be}{\begin{equation}}
\newcommand{\ee}{\end{equation}}
\newcommand{\nn}{\nonumber}
\newcommand{\bi}{\bibitem}
\newcommand{\wvp}{\bar P}

\newcommand{\ms}{\mathstrut}

\begin{document}

\title{Non-Markovian effects in strong-field pair creation }
\author{S. Schmidt, D. Blaschke, G. R\"opke}
\address{Fachbereich Physik, Universit\"{a}t Rostock,
         D-18051 Rostock, Germany}
\author{A.V. Prozorkevich, S.A. Smolyansky}
\address{Physics Department, Saratov State University, Saratov, Russian Federation}
\author{V.D. Toneev}
\address{Bogoliubov Laboratory of Theoretical Physics, Joint Institute for Nuclear Research, \\141980 Dubna, Russian Federation}
\maketitle
\begin{abstract}
We analyze a quantum kinetic equation describing  both boson and 
fermion pair production and explore analytically and numerically the solution of the non-Markovian kinetic equation. In the Markovian limit of the kinetic equation 
we find an analytical  solution for the single particle distribution function of bosons and fermions.
The numerical investigation for a homogeneous, constant electric field shows an enhancement (bosons) or a suppression (fermions) 
of the pair creation rate according to  the  symmetry character of the produced particles. For strong fields non-Markovian effects are important while they disappear for weak fields. Hence it is sufficient to apply the low density limit for weak fields but necessary to take into account memory effects for strong fields.  \\[2mm]
\end{abstract}

\section{Introduction}\label{intro}
A proper description of the pre-equilibrium evolution of the quark-gluon plasma [QGP], believed to be 
created in an ultrarelativistic heavy-ion collision can start from a transport equation  that incorporates a source term and a collision integral \cite{bialas}. The source term describes the production of pairs of particles and antiparticles while the  collisions lead to thermalization.

The formation of the QGP is assumed to proceed via the creation of a strong 
chromoelectric field in the region between the two receding nuclei after the collision. The field  subsequently decays by emitting 
quark-antiquark pairs according to the nonperturbative tunneling process of 
the Schwinger mechanism \cite{Sau,HE,Sch}. Within the flux tube model a lot of promising research has been carried out \cite{KM,GKM,nussinov,agi,matsui,Bhal,greiner,knoll,pavel,CMM}. 

The process of pair production within the Schwinger mechanism  has been addressed by many authors \cite{KM,GKM,Bhal,NN} in recent years, with the back reaction scenario also considered \cite{Back}. 
A recent application of the Schwinger mechanism of pair creation to the QCD case, 
solving a transport equation in boost invariant variables with a simple 
collision and source term, has been performed in \cite{nayak}. Another interesting calculation was provided by \cite{artur} where the polarization of the created quark-antiquark pair in a string fragmentation model for inclusive reactions was discussed.

While these studies have been very useful in exploring the preequilibrium physics of a heavy ion collision, there is still the open question how to link the field theoretical treatments with a kinetic theory. 
 More recent investigations performed for QED have shown that a consistent field theoretical approach leads to a kinetic equation with a modified  source term providing a non-Markovian evolution of the distribution function. This result was first obtained by Rau \cite{Rau} for the case of fermion pair creation for a constant electric field using a projection method \cite{zwanzig}. Therein it was emphasized that the particle pairs are produced with a non-trivial momentum dependence and not only at zero longitudinal momentum as assumed in previous studies.   

A generalized treatment allowing for a time dependent field was given in \cite{gsi,basti}. Therein we derived a kinetic equation containing both  boson and fermion pair production. For the case of a constant field our results agree with those of \cite{Rau,kme}.    
The properties of the source term itself such as the momentum dependence and the time structure have been studied \cite{basti,kme,basti1} for a constant and a time dependent field. A detailed analysis of the boson pair creation   combined with a systematic numerical study of the time structure of the solution was provided in the approach of \cite{kme}. These studies are performed for weak fields where the Markovian approximation is valid.

However, the appearence of non-Markovian aspects of the pair creation process which are expected for strong fields are not yet fully understood. Therefore the main goal of this article is the study of the time evolution of the system in a regime where memory effects become important. As an illustrative example we consider pair creation in QED.
Attention will be paid to the numerical analysis of the evolution of 
the distribution function for both bosons and fermions. We explore the influence of the different quantum statistical properties of boson and fermion pair creation and how  the differences depend on the strength of the external field.

In Section II.A we discuss   the kinetic equation with a new source term
 for pair production. In Section II.B we explore the numerical solution of the 
non-Markovian kinetic equation for weak and strong fields. We also study the influence of 
the symmetry character of fermions and bosons on the evolution of the distribution function. In Section II.C we discuss the analytic solution in the Markovian and the low density limits, respectively.  We compare the results with the non-Markovian solution and discuss the differences. The results are summarized in Section III. Within this study we neglect any influence  due to collisions as well as back reactions of the produced charged particles on the initial electric field.
 
\section{The non-Markovian kinetic equation}
\label{sec:1}
\subsection{A source term with non-Markovian character}
We consider particle production in a  strong external electric field which leads to an unstable vacuum that can decay by creation of  electron-positron pairs.
Using the field-theoretical model   of  charged particles
in an external, homogeneous, time-dependent field characterized by the vector potential $A_\mu=(0,0,0,A(t))$ with $A(t)=A_3(t)$ and 
the  resulting electric field 
$E(t)=E_3(t)=-{\dot A}\ms(t)=-dA(t)/dt$, the kinetic equation is  derived starting from the Dirac- (Klein-Gordon-) equation for fermions (bosons). The transition from the in-state to the instantaneous, quasiparticle  state at the time $t$ has been achieved by a time-dependent Bogoliubov transformation. Details of this derivation are given in  \cite{gsi,basti}. 
As the final result we obtain
the kinetic equation for the single particle distribution function 
$
f(\wvp,t)=<0|a^\dagger_{\wvp} (t)a_{\wvp} (t)|0>
$
defined as the vacuum expectation value in the time dependent basis
of creation and annihilation operators 
$
a^\dagger_{\wvp} (t), a_{\wvp} (t)
$ 
for electron states at the time $t$ and the 3- momentum $\wvp$
\bea\label{10}
&&\frac{df_\pm(\wvp,t)}{dt}=\frac{\partial f_\pm(\wvp,t)
}{\partial t}+eE(t)\frac{\partial f_\pm(\wvp,t)}{\partial P_\parallel(t)}=\\\nn
&&
\frac{1}{2}{\cal W}_\pm(t)\int_{-\infty}^t dt'{\cal W}_\pm(t') [1\pm2f_\pm(\wvp,t')]\cos[x(t',t)]\,,
\eea
where the upper sign (lower sign) in Eq. (\ref{10}) corresponds to boson (fermion) pair creation.
Details of the derivation are given in  \cite{gsi,basti,kme}. The momentum is defined as $\wvp=(p_1,p_2,P_\parallel(t))$, with the longitudinal momentum $P_\parallel(t)=p_\parallel-eA(t)$ where $p_\parallel=p_3$.
For fermion creation we find in agreement with \cite{Rau}
\be\label{12}
{\cal W}_-(t)=\frac{eE(t)\varepsilon_\perp}{\omega^2(t)}\,,
\ee
and for boson production our result agrees with \cite{kme}
\be\label{14}
{\cal W}_+(t)=\frac{eE(t)P_\parallel(t)}{\omega^2(t)}=\frac{P_\parallel(t)}{\varepsilon_\perp}{\cal W}_-(t).
\ee
We define the total energy $\omega(t)=\sqrt{\varepsilon_\perp^2+P_\parallel^2(t)}$, the transverse energy $\varepsilon_\perp=\sqrt{m^2+\vp^2_\perp}$ and the transverse momentum $\vp_\perp = (p_1,p_2)$.
Furthermore $x(t',t) = 2[\Theta(t)-\Theta(t')]$ denotes the difference of the dynamical phases which are defined as
\be
\label{30}
\Theta(t) = \int^t_{-\infty}dt'\omega(t')\,\,.
\ee
Equation (\ref{10}) is characterized by the following properties. (i) The particles are produced not only at rest $p_\parallel =0$ as assumed in more phenomenological approaches, e.g. \cite{Back}.   In Fig. 1 the dependence of the source term in low density limit for fermion production on the parallel momentum 
\be\label{45}
S^0_\pm(\wvp,t)=\frac{1}{2}{\cal W}_\pm(t)\int_{-\infty}^t dt'{\cal W}_\pm (t')\cos[x(t',t)]\,,
\ee
is shown. This calculation has been performed for a strong and weak constant electric field, $E(t) = const$. The production rate is peaked at about zero momentum, for positive momenta it approaches zero and for negative momenta it is dominated by oscillations due to the choice of a constant electric field. It increases with increasing strength of the external field. Similar results have been obtained recently by different authors \cite{Rau,gsi,basti,kme}. (ii) The source 
term and the distribution function have a momentum dependence accounted for by the  transverse energy $\varepsilon_\perp$ and by the kinetic momentum $P_\parallel(t)$, i.e. once a solution for $p_\parallel=0$ and $p_\perp=0$ is obtained, the solution for nonvanishing momenta can be generated by simple variable transformations. Therefore, we drop the explicit notation of the dependence on $\wvp$ in the distribution function and in the source term.

(iii) Furthermore, the kinetic equation (\ref{10}) has non-Markovian character due to the explicit dependence of the source term on the time evolution of the distribution function. The source term contains a time integration over the statistical factor  $[1\pm2f_\pm(\wvp,t)]$ which may cause  memory effects.  This important property will be discussed within this article. Investigating the differences of boson and fermion pair creation for weak and strong fields, we go beyond recent studies, e.g. by \cite{basti,kme}, in the following subsections. 

\subsection{Solution of the non-Markovian kinetic equation}
In the previous subsection we have discussed general features of the source term, now we want to study the time structure of the non-Markovian solution in detail by solving the kinetic equation  
\bea\label{nm}
\frac{d\,f_\pm(t)
}{d\,t}= \frac{1}{2}{\cal W}_\pm(t)\int_{-\infty}^t dt' {\cal W}_\pm(t')[1\pm2f_\pm(t')]\cos[x(t',t)]
\eea
numerically. The non-Markovian character is obvious. The appearance of  the statistical factor under the time integral means that the solution of the differential equation depends on the full time evolution of the distribution function and hence memory effects are included. This complicated structure requires a self-consistent scheme for the numerical solution.

As initial conditions we use $\lim\limits_{t\to-\infty}f_\pm(t)=0$. We solve the differential equation within the standard methods  and obtain the self-consistent solution by iteration. 
Another possibility to solve this equation is the direct solution by integration. Numerical investigations have shown that this method leads to the same result but more iteration steps are needed to find the solution and thus we decided to solve (\ref{nm}) in its differential form. 
In order to demonstrate the numerical solution of the kinetic equation we choose the simple case of  a constant electric field $E(t) = E = const.$. On the one hand this is the most simple possible case also for the numerical treatment since this assumption leads to a reduction of the number of integrations. On the other hand this is the standard {\em Ansatz} for the case that  back reactions are not included \cite{Rau,kme} and therefore permits to compare the results. The vector potential is in this case 
\be\label{Acon}
{\tilde A}(\tau) = A(\tau)/\varepsilon_\perp = -\tau E_0/e\,,
\ee
where the dimensionless variable  $E_0=eE/\varepsilon_\perp^2$ does not depend on time and the energy is given as 
\be\label{energycon}
\omega_0(\tau) =\sqrt{ 1 + E_0^2(\tau -\Delta\tau)^2}\,.
\ee
In our calculations we keep the transverse momentum fixed and hence normalize in units of a constant $\varepsilon_\perp$.  In Eqs. (\ref{Acon}) and (\ref{energycon}) we have introduced the dimensionless time variables
\be
\tau = t\,\varepsilon_\perp\,,
\ee
and
\be
\Delta\tau=\Delta\tau(p_\parallel,\vp_\perp)=\frac{p_\parallel\varepsilon_\perp(\vp_\perp)}{eE}\,.
\ee
This notation is also convenient to distinguish the weak field ($E_0<1$) and strong field ($E_0>1$) limits. 

For such a simple model case, we can solve the integral of the dynamical phase (\ref{30}) and obtain
\bea
\Theta(\tau)
=\frac{1}{2}(\tau-\Delta\tau)\omega_0(\tau)
+\frac{1}{2E_0}\ln\bigg[E_0(\tau-\Delta\tau)+\omega_0(\tau)\bigg]\,.
\eea
In the following we discuss the numerical solution of the kinetic equation for bosons and fermions for different field strengths. In Fig. 2 we plot the distribution function for bosons for weak fields and strong fields. 
The particles are produced at about zero kinetic momentum. Because of the fact that we have no damping mechanism the distribution oscillates around  a constant value. The oscillations to be seen in the figure are due to the choice of a constant electric field. In response to strong fields the frequencies of these oscillations increase while the amplitudes decrease compared with the limit at large times. Other {\em Ans\"atze} for the time dependence of the external field such as a Gaussian shape \cite{basti} may lead to a damping of these oscillations. But it is important to note that in order to describe a more realistic situation the back reaction of the produced particles on the initial field should be included \cite{Back,kme}. 
The curves are normalized to the large time limit 
\be
f(\tau\rightarrow\infty)=\exp\bigg(\frac{-\pi}{E_0}\bigg)\,.
\ee
We observe that for all plotted field strengths the curves converge to this limit. With other words one can conclude that for large times we obtain the old result given by Schwingers formula for weak fields as well as for strong fields. The absolute value of the  distribution function at $\tau\rightarrow\infty$ is larger for strong fields than it is for weak fields. 

We obtain a similar result for fermion pair creation, see Fig. 3. Due to the different amplitudes for fermion and boson pair production, Eqs. (\ref{12}) and (\ref{14}), the shapes of the curves are slightly different, in particular around zero kinetic momenta. We can compare the distribution function for fermions and bosons for a given  field strength, Fig. 2 and Fig. 3. The onset  of particle creation (first maximum) is earlier for fermions than it is for bosons. Both curves reach the same limit for large times and oscillate  with the same frequency for a given $E_0$. The results which we obtain for the boson case are in agreement with those of \cite{kme} for the case of weak fields, wherein the Markovian limit was employed. But the  numerical solution introduced in this  section allows in addition to consider pair creation for strong fields.

\subsection{The low density and the Markovian limit}
\label{sec:2}
\subsubsection{ A closed kinetic equation}
In the previous subsection we have demonstrated how to solve the kinetic equation in its non-Markovian form and explored the numerical solutions. In this Section we want to discuss the solution of the kinetic equation (\ref{10}) for 
weak fields where it is possible to apply approximations. These are the Markovian approximation and the low density limit \cite{basti1}.

The Markovian limit of the non-Markovian equation  Eq. (\ref{10}) is defined by the neglect of memory effects in the source term. Technically this means that we approximate the distribution function occuring under the integration by $f(\tau')\rightarrow f(\tau)$. We obtain the following kinetic equation
\be
\label{40}
 \frac{d\,f^M_\pm(\tau)}{d\tau}= [1\pm
2f^M_\pm(\tau)]S^0_\pm(\tau)=S^M_\pm(\tau),   
\ee
where $S^0_\pm(\tau)$ is the source term in low density limit, Eq. (\ref{45}).

Taking into account the initial condition
$\lim\limits_{\tau\to-\infty}f^M_\pm(\tau)=0$, we obtain the following solution of the kinetic equation (\ref{40})
\be
\label{50}
f^M_\pm(\tau)=\mp\frac{1}{2} \bigg(1-\exp\bigg[\pm 2\int_{-\infty}^\tau d\tau'S^0_\pm(\tau')\bigg]\bigg)\,.
\ee
This result is exact in the Markovian limit and holds for any time-dependent homogeneous electric field.  In the lowest order of the expansion of Eq. (\ref{50}) we obtain
\be
\label{55}
f^0_\pm(\tau) = \int_{-\infty}^\tau d\tau'S^0_\pm(\tau')\,.
\ee
This solution is equivalent to the low density limit where effects due to the symmetry character of the created particles are neglected in the source term. Note that both the low density limit and the Markovian approximation are restricted to weak fields, $E_0<1$.

Before we discuss the numerical results for the Markovian approximation and the low density limit, we explore  general properties of the distribution function defined in Eq. (\ref{50}).
One requirement of a kinetic theory is that the distribution function for bosons and for fermions must be positive definite for all times and momenta.
In order to prove the validity of this important property we rewrite Eq. (\ref{55}) as 
\bea\label{60}
f^0_\pm(\tau)=\frac{1}{2} \int\limits ^\tau_{-\infty}d\tau'g^1_\pm(\tau')
\int\limits ^{\tau'}_{-\infty}d\tau''g^1_\pm(\tau'')
+ \frac{1}{2} \int\limits
^t_{-\infty}d\tau' g^2_\pm(\tau')\int\limits ^{\tau'}_{-\infty}d\tau''g^2_\pm(\tau'')\,,
\eea
 where the functions $g^{1,2}_\pm(\tau)$ are given as
\be
g^{1,2}_\pm(\tau)={\cal W}_\pm(\tau)
\left\{\begin{array}{c}\cos[2\Theta (\tau)]\\ \sin[2\Theta
(\tau)]\end{array}\right\}.
\ee
It is easy to transform the integral ({\ref{60}) to the
following form
\be
\label{70}
f^0_\pm(\tau)=\frac{1}{4} \bigg(\int\limits
^\tau_{-\infty}d\tau'g^1_\pm(\tau')\bigg)^2 + \frac{1}{4}\bigg(\int\limits
^\tau_{-\infty}d\tau'g^2_\pm(\tau')\bigg)^2 \ . 
\ee
From this quadratic representation of Eq. (\ref{55}) we can conclude that the distribution function is positive definite as it should be,  
\be
f^0_\pm(\tau)\ge 0 \ .
\ee
On the other hand this result is very useful for numerical calculations since Eq. (\ref{70}) just requires to perform a single integration compared with Eq. (\ref{55}) in which it was necessary to solve a double integration, hence it makes the numerical treatment  easier. 

At large times, $\tau\rightarrow\infty$, the distribution function is in the asymptotic regime. In low density limit it is easy to show the feature that $f_0(\infty)\neq 0$. The distribution function evolves from zero at $\tau\rightarrow-\infty$ to an asymptotic nonzero value. In the absence of back reaction and collisions (or any other damping mechanism) we observe an accumulation effect. In the next subsection we will elucidate these properties with numerical results.  

\subsubsection{Numerical results}\label{sec:3}

Using the results of Eqs. (\ref{70})  we can explore the solution of Eq. (\ref{50}) for the distribution functions  in the Markovian limit and in the low density approximation.  

In Figs. 4 and 5 we compare the distribution functions of the non-Markovian solutions with the low density limit solutions for fermions and bosons, respectively. In the lower panels we have chosen a weak field and observe that the results agree with each other. This was expected since for weak fields the absolute value of the distribution function is small and consequently the statistical factor $[1\pm 2f(\tau)]$ does not considerably deviate from one.  Hence the low density limit is a good approxmation for relatively small field strengths, $E_0<1$.

In the upper panels we have chosen large field strengths. Both for fermions, Fig. 4, and for bosons, Fig. 5, the low density limit solution shows a different limit for large times. The inclusion of the quantum statistical character in the non-Markovian solution leads for fermion pair creation to a suppression and for boson pair creation to an enhancement compared to the corresponding low density limit. These plots elucidate two important things. On the one hand they demonstrate the influence of the different symmetry character of fermions and bosons. On the other hand we see that for strong fields the low density limit solution provides wrong results. Technically the reason is clear. The statistical factor $[1\pm 2f(\tau)]$ deviates from one and therefore it has to be included into the kinetic equation. Physically this means that the non-Markovian character becomes important.

We have to distinguish different time scales: the memory time and the production time \cite{Rau,basti,kme,basti1}. While the memory time 
has quantum mechanical origin and can be considered as the time needed to tunnel the barrier, the production time is the time interval between two creation processes. The memory time is now for strong fields of the same order of magnitude as the production time. A separation of the time scales, which is necessary for the approximation discussed in the previous subsection, is no longer possible. The pre-history affects the evolution of the distribution function and therefore memory effects become important. This is taken into account by the time integration over the statistical factor in the full non-Markovian equation.

In the previous subsection we have also discussed the Markovian approximation. It is characterized by the neglect of memory, but the different statistical character of fermions and bosons is still taken into account. In Fig. 6 we compare the three solutions of the kinetic equation for strong fields for bosons (lower panel) and for fermions (upper panel). Besides the already discussed non-Markovian solution and the low density approximation, we also show the Markovian limit. Needless to point out that for weak fields the Markovian solution also agrees with the non-Markovian solution. Although the Markovian solution is in better agreement with the correct solution than the low density limit, the error is visible and the Markovian limit fails for strong fields.  For weak fields the low density limit provides exact results since non-Markovian effects disappear. For strong fields it is unavoidable to solve the non-Markovian equation. Just in a small band of field stengths of about one the Markovian limit is a sensible approximation, better than the low density limit but non-Markovian effects are still very small. Hence for small fields, $E_0 < 1$, we suggest to use the low density limit, for strong fields, $E_0 > 1$, the non-Markovian solution.  

\section{Summary}\label{summary}
We have analytically and numerically explored the solution of a quantum kinetic equation describing particle production of boson and fermion pairs. The source term providing the creation of pairs is characterized by  its non-Markovian character. The time evolution of the distribution function depends on the entire pre-history of the evolution.
We have numerically solved  the kinetic equation in its non-Markovian form  for both weak and strong fields. We compared these solutions with both the low density limit and the Markovian approximation
for which we found an analytic solution in closed form. 

We observe that for strong fields the results depend on the statistical character of the produced particles.  We obtain a suppression of the production rate for the creation of fermion pairs and an enhancement for boson pairs when we compare the full solution with the approximation where memory effects are neglected.  In the domain of weak fields the exact solution of the Markovian limit can be approximated by its low density limit being the lowest order term of an expansion of the full Markovian solution. As a numerical result we obtain that indeed for weak fields the use of the Markovian limit as well as of the low density limit is justified. The curves calculated in the Markovian limit, the low density limit and the full solution are in good agreement. This means that non-Markovian effects disappear for weak fields. Therefore it is sufficient to use for weak fields just the low density limit.
 
Furthermore we find that for strong field strengths the Markovian approximation as well as the low density limit break down since memory effects become important. For strong fields the distribution function has a large value and therefore the statistical factor $[1\pm 2f(\tau)]$ deviates from one. Hence it is transparent that the non-Markovian character appears for strong fields. This observation indicates that it is necessary to include memory in order  to obtain e.g. the correct limit for large times.

In conclusion we can summarize that for weak fields it is appropriate to use the low density limit while for strong fields it is necessary to solve the non-Markovian kinetic equation.
In order to give a more complete picture of the physics beyond the Markovian limit it is necessary to study in a next possible step the momentum dependence of the distribution functions. 

Although the numerical analysis was performed for a constant field our results qualitatively hold for any time-dependent field. This will be important when a more realistic scenario is addressed where the electric field is determined self-consistently. In order to incorporate back reactions it is necessary to solve the Maxwell equation that determines the electric field via the conduction current and the polarization current due to the creation of the charged particle pairs \cite{Back,kme}. Furthermore, it would be of great interest to extend this approach to the QCD case in order to explore the consequences of the new source term for the pre-equilibrium physics in ultrarelativistic heavy-ion collisions. 

\acknowledgments
The authors dedicate this article to the memory of J.M. Eisenberg. One of us 
(S.S.) had the pleasure to work as a Minerva fellow at the Tel Aviv University
 with J.M.E. who has attracted him to explore this interesting field of 
physics.
S.S. acknowledges valuable discussions with C.D. Roberts and B. Svetitsky and thanks the 
Physics Division at Argonne National Laboratory as well as the nuclear physics group of the Tel Aviv University for their hospitality and 
support during visits where part of this work was conducted. 
This work was supported in
part by the State Committee of Russian Federation for Higher Education under grant N 29.15.15, by BMBF under the
program of scientific-technological collaboration (WTZ project
RUS-656-96)  and by the Hochschulsonderprogramm (HSP III) under the project No. 0037-6003.

\newpage
\begin{figure}
\caption{The dependence of the source term in low density limit on the parallel momentum for fermion pair creation for a strong field (upper panel: $E_0=1.5$) and a weak field (lower panel: $E_0=0.5$).}
\end{figure}

\begin{figure}
\caption{The boson distribution function as a solution of the non-Markovian equation as function of time for different field strengths at $p_\parallel =0$.}
\end{figure}

\begin{figure}
\caption{Same as Fig. 2 for fermions.}
\end{figure}

\begin{figure}
\caption{The time evolution of the distribution functions of bosons ($f_+$) compared with the low density limit $f^0_+$   at $p_\parallel=0$ for a strong field (upper panel: $E_0=5.0$), and a weak field (lower panel: $E_0=0.7$), is shown.}
\end{figure}

\begin{figure}
\caption{The time evolution of the distribution functions of fermions ($f_-$) compared with the low density limit $f^0_-$   at $p_\parallel=0$ for a strong field (upper panel: $E_0=3.0$), and a weak field (lower panel: $E_0=0.5$), is shown. }
\end{figure}

\begin{figure}
\caption{The time evolution of the distribution functions of fermions (upper panel: $E_0=3.0$) and bosons (lower panel: $E_0=5.0$) within different approximations for strong fields. Only the full non-Markovian solution provides the correct limit for large times.}
\end{figure}

\newpage

\centering{
\epsfig{figure=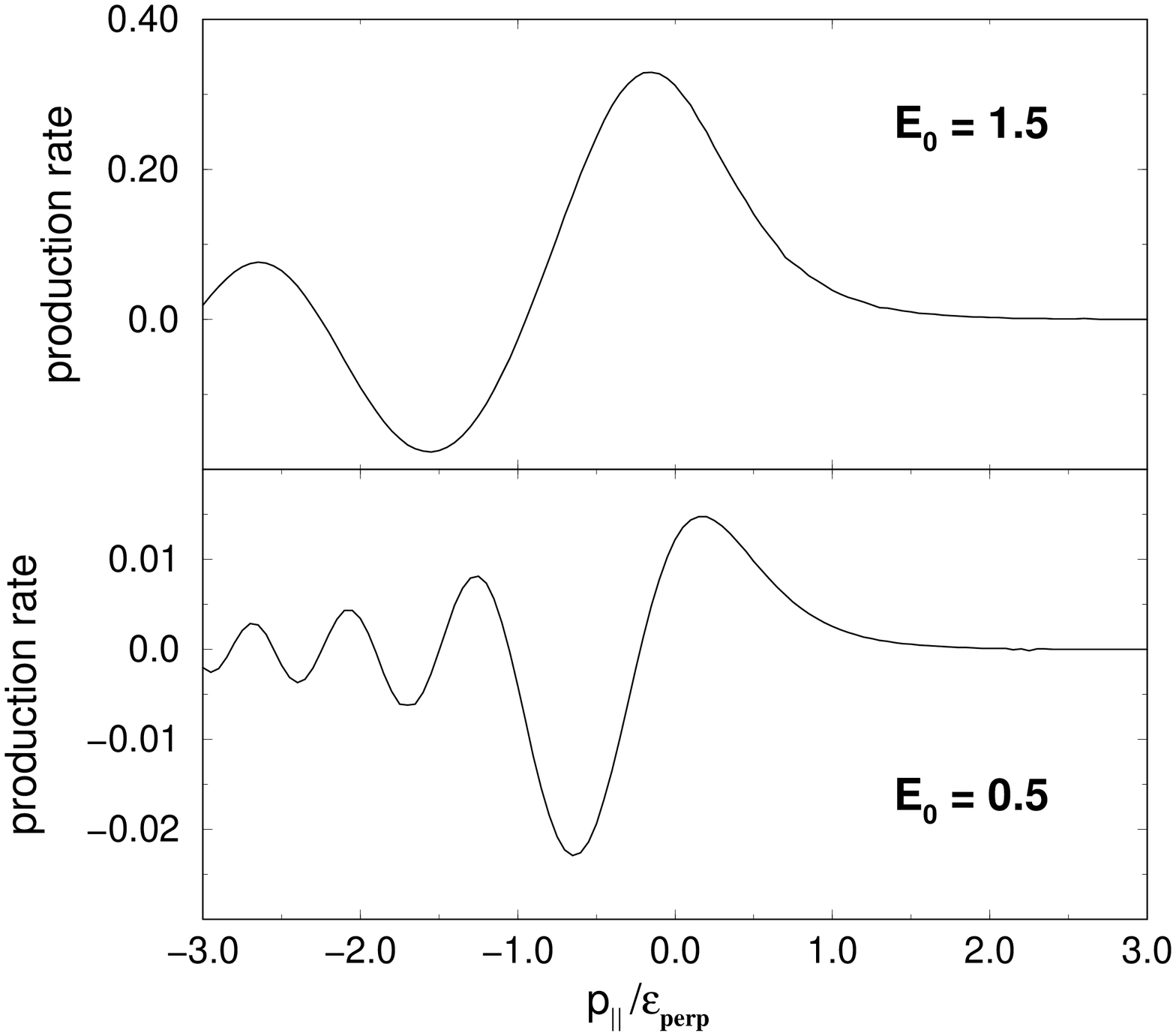,width=13cm,height=13cm,angle=-90}}

\vspace{2cm}
{\LARGE FIG. 1}

\newpage

\centering{
\epsfig{figure=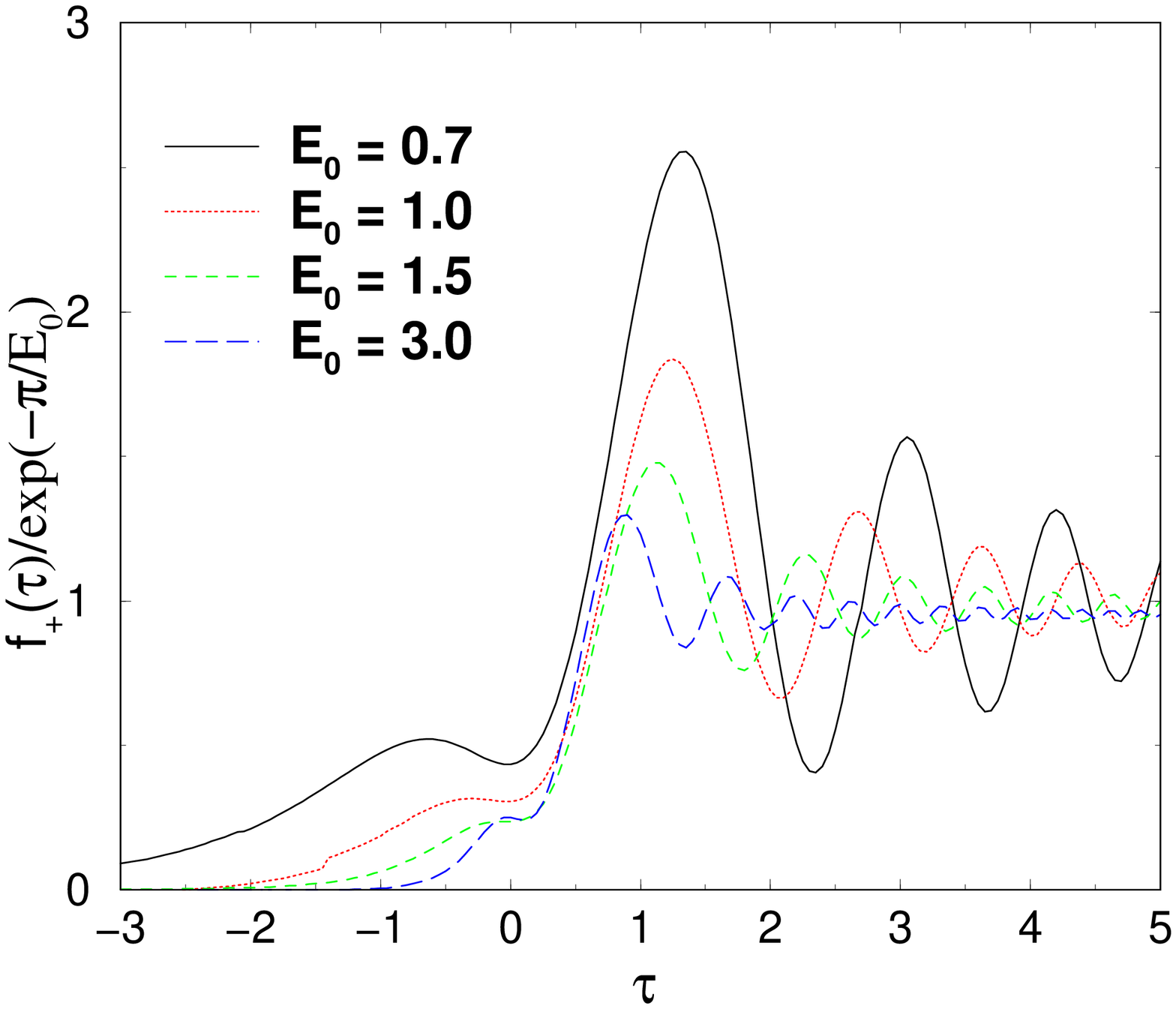,width=13cm,height=13cm,angle=-90}}

\vspace{2cm}
{\LARGE FIG. 2}

\newpage
\centering{
\epsfig{figure=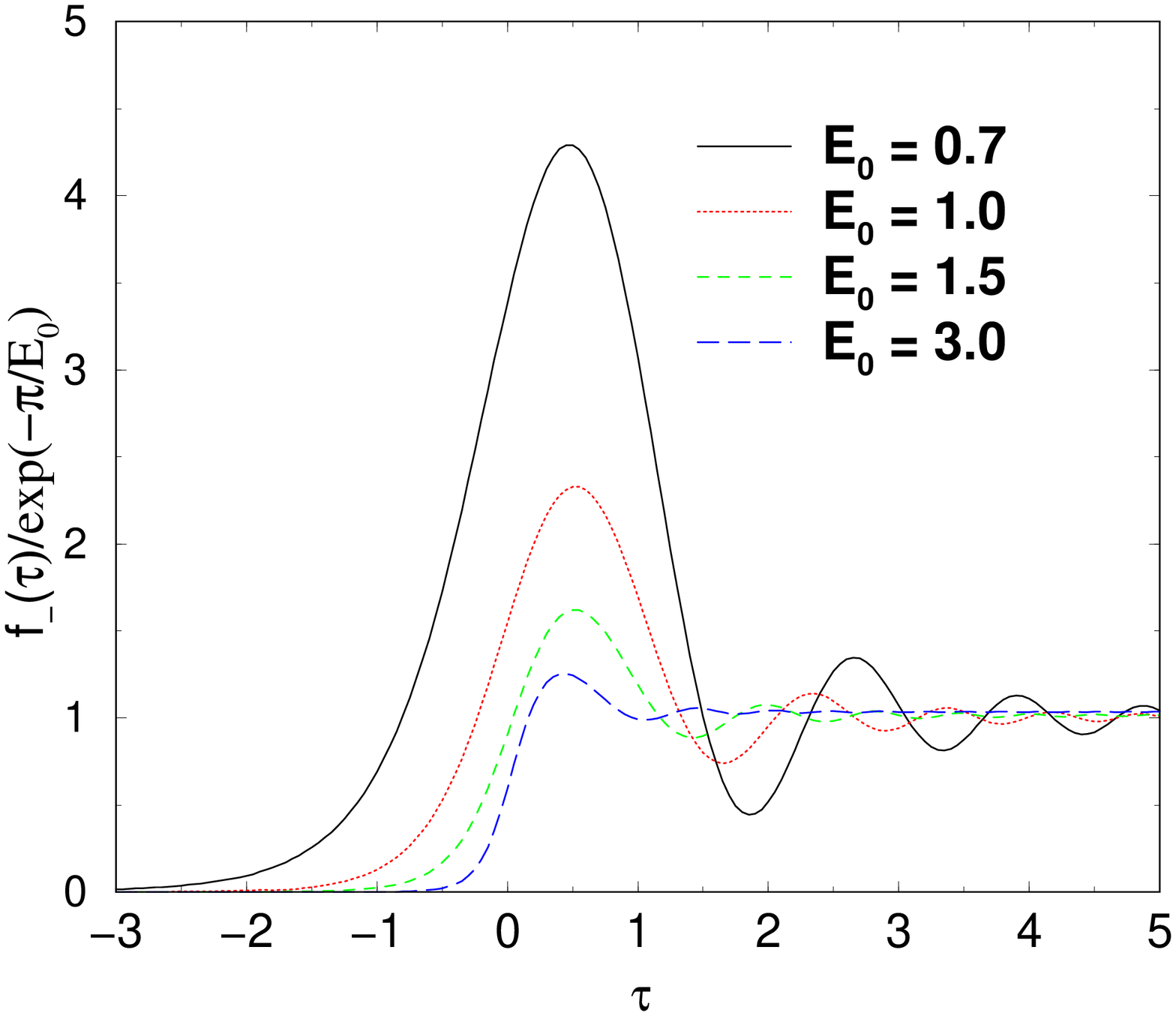,width=13cm,height=13cm,angle=-90}}

\vspace{2cm}
{\LARGE FIG. 3}

\newpage
\centering{
\epsfig{figure=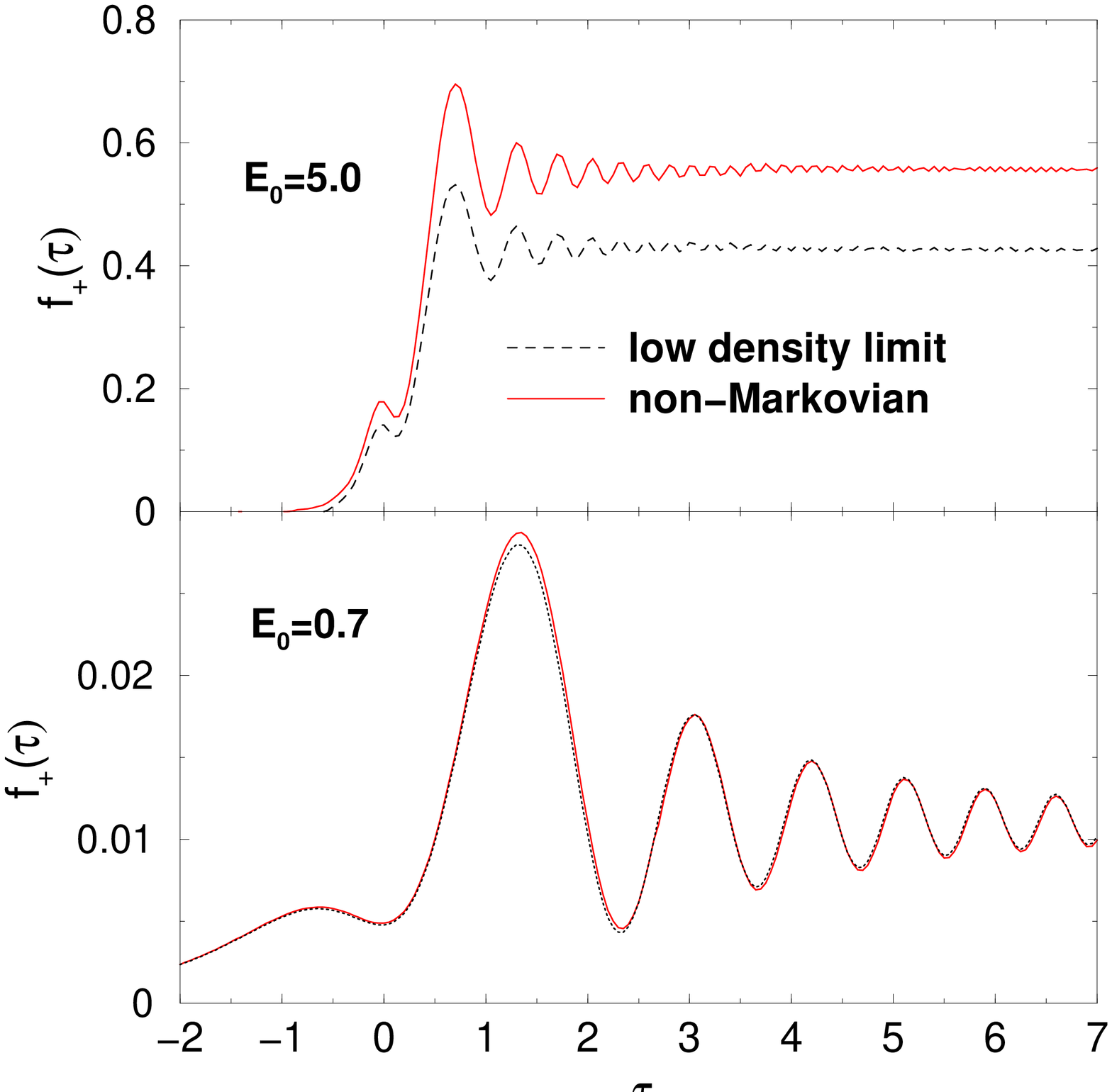,width=15cm,height=12cm,angle=-90}}

\vspace{2cm}
{\LARGE FIG. 4}

\newpage
\centering{
\epsfig{figure=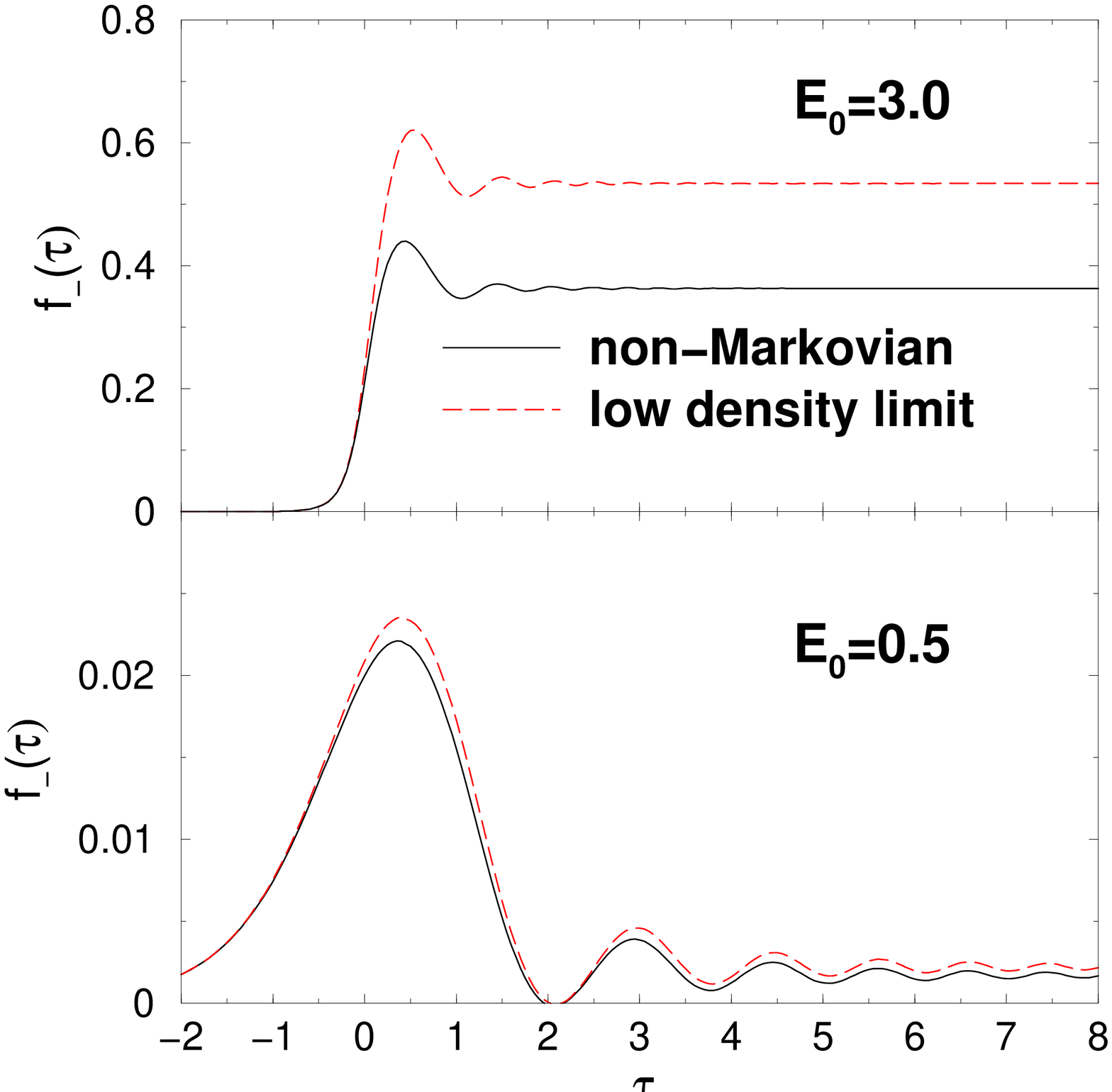,width=15cm,height=12cm,angle=-90}}

\vspace{2cm}
{\LARGE FIG. 5}

\newpage
\centering{
\epsfig{figure=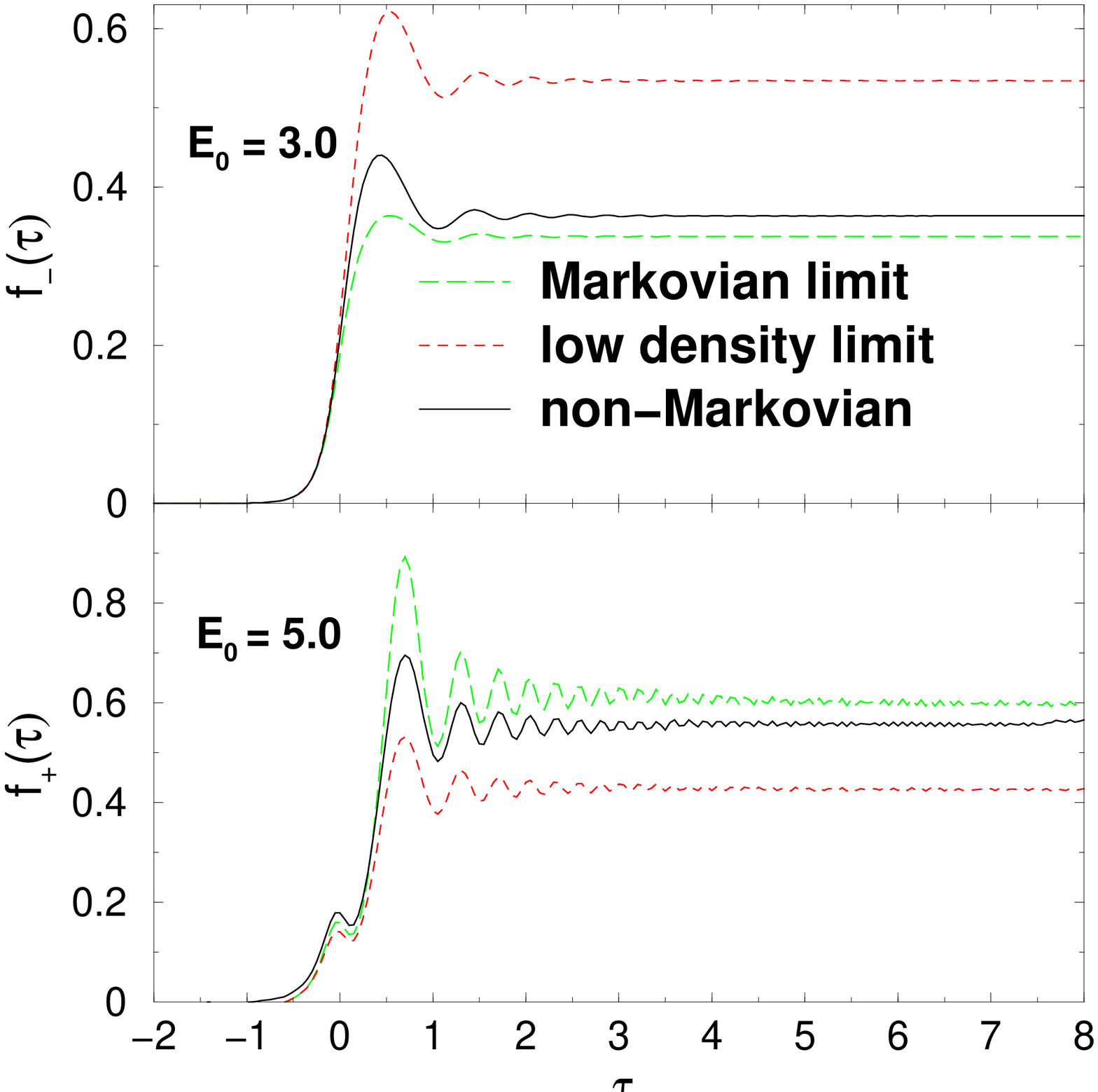,width=15cm,height=12cm,angle=-90}}

\vspace{2cm}
{\LARGE FIG. 6}


\begin{references}
\bi{bialas}{A.~Bia\l as and W.~Czy\.z, Phys.~Rev.~{\bf  D 30} (1984) 2371;
{\bf 31} (1985) 198; Z.~Phys.~ {\bf C 28} (1985) 255; Nucl.~Phys.~{\bf B
267} (1985) 242; Acta Phys.~Pol.~{\bf B 17} (1986) 635.}
\bi{Sau} {F.~Sauter, Z.~Phys.~{\bf 69} (1931) 742.}
\bi{HE} {W.~Heisenberg and H.~Euler, Z.~Phys.~{\bf 98} (1936) 714.}
\bi{Sch} {J.~Schwinger, Phys.~Rev.~{\bf 82} (1951) 664.}
\bi{KM} {K.~Kajantie and T.~Matsui, Phys. Lett. {\bf B 146} (1985) 373.}
\bi{GKM}{ G.~Gatoff, A.K.~Kerman, and T.~Matsui, Phys.~Rev.~{\bf D 36} (1987) 114.}
\bi{nussinov}{S. Nussinov, Phys. Rev. Lett. {\bf 34} (1975) 1286.}
\bi{agi}{B. Andersson et al., Phys. Rep. {\bf 97} (1993) 31.}
\bi{matsui}{N.K. Glendenning and T. Matsui, Phys. Rev. {\bf D 28} (1983) 2890.}
\bibitem{Bhal} {R.S. Bhalerao and V. Ravishankar, Phys. Lett.  {\bf B 409} (1997)
38.}
\bi{greiner} {W. Greiner, B. M\"uller, and J. Rafelski, {\em Quantum Electrodynamics of Strong Fields} (Springer-Verlag, Berlin, 1985).}
\bi{knoll}{M. Herrmann and J. Knoll, Phys. Lett.  {\bf B 234} (1990) 437.} 
\bi{pavel}{H.-P. Pavel and D.M. Brink, Z. Phys.  {\bf C 51} (1991) 119.}
\bi{CMM} {A.A. Grib, S.G. Mamaev and V.M. Mostepanenko,
{\em Vacuum quantum effects in strong external fields}, (Atomizdat,
Moscow, 1988).}
\bi{NN} {N.B. Naroshni and A.I. Nikishov,
Yad. Fiz. {\bf 11} (1970) 1072 (Sov. J. Nucl. Phys. {\bf 11} (1970) 596);
V.S. Popov and M.S. Marinov, Yad. Fiz. {\bf 16} (1972) 809 (Sov. J. Nucl. Phys. {\bf 16} (1974) 449);
V.S. Popov, Zh. Eksp. Teor. Fiz. {\bf 62}
(1972) 1248 (Sov. Phys. JETP {\bf 35} (1972) 659);
M.S. Marinov and V.S. Popov, Fortsch.   Phys. {\bf 25}
(1977) 373.}
\bi{Back} {Y.~Kluger et al., Phys.~Rev.~Lett.~{\bf 67} (1991) 2427;\\F.~Cooper et al., Phys.~Rev. {\bf D 48} (1993) 190;\\
Y.~Kluger, J.M.~Eisenberg, and B.~Svetitsky,
Int.~J.~Mod.~Phys. {\bf E 2} (1993) 333 (herein further references
 may be found).}
\bi{nayak}{ G.C. Nayak and V. Ravishankar, Phys. Rev.  {\bf C 58} (1998) 356; Phys. Rev.  {\bf D 55} (1997) 6877.}
\bi{artur}{X. Artru and J. Czyzewski, Acta Phys. Pol. {\bf B 29} (1998) 2115.}
\bi{Rau} {J. Rau, Phys. Rev.  {\bf D 50} (1994) 6911.}
\bi{zwanzig}{R. Zwanzig, Physica {\bf 30} (1964) 1109.}
\bi{gsi}{S.A. Smolyansky et al., {\em Dynamical derivation of a quantum kinetic equation for particle production in the Schwinger mechanism}, e-print archive, hep-ph/9712377, GSI-Preprint-97-72, December 1997.}
\bi{basti}{S. Schmidt et al., {\em A quantum kinetic equation for particle 
production in the Schwinger mechanism}, Int.~J.~Mod.~Phys.~E, in press, hep-ph/9809227 .}
\bi{kme}{Y. Kluger, E. Mottola, and J.M. Eisenberg, {\em The quantum Vlasov equation and its Markov limit}, Phys. Rev. {\bf D}, in press, hep-ph/9803372.}
\bi{basti1}{S. Schmidt, A.V. Prozorkevich, S.A. Smolyansky, 
{\em Creation of boson and fermion pairs in strong fields}, 
Proceedings 'V. Workshop on Nonequilibrium Physics at Short Time Scales', April
27-30, 1998; hep-ph/9809233. }
\end{references}
\end{document}